%Paper: hep-ph/9308292
%From: anelson@sphal.UCSD.EDU (Ann E. Nelson)
%Date: Wed, 18 Aug 1993 12:00:23 -0700

\input harvmac
%\draftmode
%%%%%%%%%%%%%%%%%%%%%%%%%%%%%%%%%%%%%%%%%%%%%%%%%%%%%%%%%%%%%%%%%%%%%%
%
%  UCSD macros to overwrite some of the definitions in harvmac.tex
%  (include after harvmac.tex)
%  last modified 4/92
%
%%%%%%%%%%%%%%%%%%%%%%%%%%%%%%%%%%%%%%%%%%%%%%%%%%%%%%%%%%%%%%%%%%%%%%%
%
% modify the output routine for the little format
%
\ifx\answ\bigans
\else
\output={
  \almostshipout{\leftline{\vbox{\pagebody\makefootline}}}\advancepageno
}
\fi
%
%
% address
%
\def\mayer{\vbox{\sl\centerline{Department of Physics 0319}%
\centerline{University of California, San Diego}
\centerline{9500 Gilman Drive}
\centerline{La Jolla, CA 92093-0319}}}
\def\title#1{\nopagenumbers\hsize=\hsbody%
\centerline{\titlefont #1} \tenpoint \vskip .5in\pageno=0}
%
%
% grant numbers
%
\def\doe{\#DOE-FG03-90ER40546}
\def\tnlrc{\#RGFY93-206}

\def\pyidk{PHY-9057135}
%
% preprint number
%
\def\UCSD#1#2{\noindent#1\hfill #2%
\bigskip\supereject\global\hsize=\hsbody%
\footline={\hss\tenrm\folio\hss}}% restores pagenumbers
%
% abstract
%
\def\abstract#1{\centerline{\bf Abstract}\nobreak\medskip\nobreak\par #1}
%
%
% titlefont
%
%
\edef\tfontsize{ scaled\magstep3}
 \tfontsize  \tfontsize
 \tfontsize \font\titlei=cmmi10 \tfontsize
\font\titleis=cmmi7 \tfontsize \font\titleiss=cmmi5 \tfontsize
\font\titlesy=cmsy10 \tfontsize \font\titlesys=cmsy7 \tfontsize
\font\titlesyss=cmsy5 \tfontsize  \tfontsize
\skewchar\titlei='177 \skewchar\titleis='177 \skewchar\titleiss='177
\skewchar\titlesy='60 \skewchar\titlesys='60 \skewchar\titlesyss='60
%
%\def\titlefont{\def\rm{\fam0\titlerm}% switch to title font
%\textfont0=\titlerm \scriptfont0=\titlerms \scriptscriptfont0=\titlermss
%\textfont1=\titlei \scriptfont1=\titleis \scriptscriptfont1=\titleiss
%\textfont2=\titlesy \scriptfont2=\titlesys \scriptscriptfont2=\titlesyss
%\textfont\itfam=\titleit \def\it{\fam\itfam\titleit}\rm}
%
%
% math symbols
%
%---------------------------------------------------------------------
%
%
%
% space and backspace in l mode
%
\def\lspace{\ifx\answ\bigans{}\else\qquad\fi}
\def\lbspace{\ifx\answ\bigans{}\else\hskip-.2in\fi} % $$\lbspace...$$
%
%
%     curly letters
%
   %curly letters

  \def\CO{{\cal O}}

%
%
%
%     derivatives
%
%

%

\def\bar#1{\overline{#1}}

%
%
 %puts a small half in a displayed eqn
\def\frac#1#2{{\textstyle{#1\over #2}}} %puts a small fraction
%in a displayed eqn
%
%
%     various math operators
%
%

\def\GeV{{\rm GeV}}
\def\MeV{{\rm MeV}}

\def\eV{{\rm eV}}
\def\TeV{{\rm TeV}}
%
%
%
%       relations
%
\def\ltap{\ \raise.3ex\hbox{$<$\kern-.75em\lower1ex\hbox{$\sim$}}\ }
\def\gtap{\ \raise.3ex\hbox{$>$\kern-.75em\lower1ex\hbox{$\sim$}}\ }
\def\gl{\ \raise.5ex\hbox{$>$}\kern-.8em\lower.5ex\hbox{$<$}\ }
\def\roughly#1{\raise.3ex\hbox{$#1$\kern-.75em\lower1ex\hbox{$\sim$}}}
%
%
%       This defines et al., i.e., e.g., cf., etc.
\def\ie{\hbox{\it i.e.}}        
\def\eg{\hbox{\it e.g.}}        
\def\etal{\hbox{\it et al.}}
\def\np#1#2#3{{Nucl. Phys. } B{#1} (#2) #3}
\def\pl#1#2#3{{Phys. Lett. } {#1}B (#2) #3}
\def\prl#1#2#3{{Phys. Rev. Lett. } {#1} (#2) #3}
\def\physrev#1#2#3{{Phys. Rev. } {#1} (#2) #3}

\def\prep#1#2#3{{Phys. Rep. } {#1} (#2) #3}

\def\frac#1#2{{\textstyle{#1 \over #2}}}

\def\[{\left[}
\def\]{\right]}
\def\({\left(}
\def\){\right)}

\hyphenation{higgs-ino}
\def\mayer{\vbox{\sl\centerline{Department of Physics}
\centerline{9500 Gilman Drive 0319}
\centerline{University of California, San Diego}
\centerline{La Jolla, CA 92093-0319}}}
{\titlefont{\centerline{Cosmological
Implications of} \centerline{Dynamical Supersymmetry
Breaking\footnote{$^*$}{This
work is supported in part by funds provided   by the U. S. Department of Energy
(D.O.E.) under contracts     \doe and DE-FG0590ER40559,  by the National
Science
Foundation under grant \pyidk, by the Sloan Foundation
and  by the Texas National Research Laboratory Commission
under grant \tnlrc. }}}}
\def\pyi{National Science Foundation Young Investigator Award.\hfill}
\def\sloan{Alfred~P.~Sloan Foundation Research Fellowship.\hfill}

\vfill
\centerline{Tom Banks}
\medskip
\vbox{\sl\centerline{ Department of Physics and Astronomy}
\centerline{\sl Rutgers University}
\centerline{\sl Piscataway, NJ 08855-0849}}
\medskip
\centerline{ David B. Kaplan
\footnote{$^\dagger$}{\pyi}\footnote{$^\#$}{\sloan}and Ann E. Nelson {$^\#$}}
\medskip \mayer
\vfill
\abstract{We provide a taxonomy of dynamical supersymmetry breaking theories,
and
discuss the cosmological implications of the various types of models.
Models in which supersymmetry breaking is produced by chiral superfields
which only have interactions of gravitational strength (\eg\ string theory
moduli) are inconsistent with standard big bang nucleosynthesis unless the
gravitino mass is greater than $\CO(3) \times 10^4$ GeV. This problem cannot be
solved by inflation. Models in which supersymmetry is dynamically broken by
renormalizable interactions in flat space have no such cosmological problems.
Supersymmetry can be broken either in a hidden or the visible sector. However
hidden sector models  suffer from several naturalness problems and have
difficulties in producing an acceptably large gluino mass.  }\vfill
\UCSD{\hbox{hep-ph@xxx/9308292, UCSD/PTH 93-26, RU-37}}{July 1993}

\newsec{\bf Introduction}

 From a practical point of view, the primary allure of supersymmetric
field theories lies in their promise to solve the gauge hierarchy
problem.  Supersymmetry succeeds in explaining why $M_W/M_P$ is not $\CO(1)$
by eliminating all quadratic divergences. However, an explanation for
the observed value $M_W/M_P = 10^{-17}$ cannot be furnished by
supersymmetry itself, but only by some theory for dynamical supersymmetry
breaking, in which the scale of the weak interactions arises from
the Planck scale and dimensionless coupling constants through
dimensional transmutation\nref\witten{E.
Witten,
\np{188}{1981}{513}}\nref\sqcd{
I. Affleck, M. Dine and
N. Seiberg,
\np{241}{1984}{493};
G.C. Rossi, G.
Veneziano,
\pl{138}{1984}{195}}\nref\susybreak{I. Affleck, M. Dine and
N. Seiberg, \prl{52}{1984}{1677}; Y. Meurice and G.
Veneziano, \pl{141}{1984}{69}}\nref\ads{I. Affleck, M. Dine and N.
Seiberg, \np{256}{1985}{557}}\refs{\witten-\ads}. In   dynamical
supersymmetry (SUSY) breakdown scenarios, SUSY is unbroken to all
orders in perturbation theory but broken by nonperturbative effects
proportional to $ e^{-a/g^2}$, where $a$ is a number of order $4 \pi^2$
and $g$ is a gauge coupling constant.  At present, there does not exist a
unique compelling scenario for dynamical supersymmetry breaking  which is
compatible with standard particle phenomenology,  and so our present
understanding of the supersymmetric resolution of the hierarchy problem
is incomplete. The purpose of the present paper is not to resolve this
problem, but to delineate the options for dynamical supersymmetry
breaking and to show that there are strong cosmological constraints on
  scenarios in which supersymmetry breaking vanishes
in the   limit in which the Planck mass is taken to infinity
with some ``intermediate'' scale fixed.  For instance, all scenarios
based on gaugino condensation in a hidden sector \ref\gauginocond{Recent
reviews include D. Amati \etal, \prep{162}{1988}{169}; H.-P. Nilles,
Int. J. Mod. Phys. A5 (1990) 4199; J. Louis, in Proceedings of the 1991
DPF meeting in Vancouver, (World Scientific, 1992)}   are of this type,
as are all models where supersymmetry breaking is driven by F terms of
string theory moduli.

We wish to provide a taxonomy of dynamical SUSY breaking models
in which the only explicit mass scale is the Planck scale.  They must
explain the $M_W/M_P$ hierarchy, and be consistent with
low energy phenomenology (including a small
or vanishing cosmological constant, achieved through fine-tuning).
The task is complicated by there being few completely generic features
to analyze.   The only feature apparently common to all models of
dynamical SUSY breaking is  the existence of an R symmetry which is either
exact or nearly exact just above the scale of SUSY breaking, and which is
spontaneously broken.  This is not a theorem, but
no known counter-examples exist \ads, and there does exist an argument that
generic
theories without a continuous R symmetry will not dynamically break
supersymmetry
\ref\seiberg{A.E. Nelson and N. Seiberg, in preparation.}. If R symmetry is
exact, however,  then it is impossible to add a constant to the superpotential
to
fine tune the cosmological constant. Thus viable dynamical SUSY breaking models
must have a spontaneously broken, approximate R symmetry.  This results in a
pseudo
Nambu-Goldstone boson---the R-axion---which  can cause phenomenological or
cosmological  problems which must be addressed.

To proceed further it is useful to distinguish between
Visible Sector (VS) models of dynamical SUSY breaking---in which
Planck scale physics plays no
role---and Hidden Sector (HS) models which make use of nonrenormalizable
operators suppressed by powers of $M_P$. In VS models
\ref\dinelson{M. Dine and A.E. Nelson,\physrev{48}{1993}{1277}.} the fields
whose
auxiliary components are responsible for the primary breaking of SUSY have
renormalizable interactions with the fields of the standard model.  In
extant examples the SUSY breaking sector
contains fields which transform nontrivially under $SU(3)\times
SU(2)\times U(1)$, and these gauge symmetries communicate SUSY breaking
to the low energy world.  Such models have several attractive features.
They do not depend on Planck scale physics, and, modulo computational
difficulties involved in solving the strongly coupled SUSY breaking sector,
they give complete predictions for the parameters of the supersymmetric
standard model (SSM).  In particular, since SUSY breaking is
communicated to the standard model via gauge interactions, squarks are
automatically nearly degenerate, and there are no problems with flavor
changing neutral currents.  Finally, gauginos have relatively large,
phenomenologically acceptable masses.

Known viable VS models have a rather complicated structure which is
necessary to give the unwanted R-axion a sufficiently large mass
to be compatible with bounds from experiment and astrophysical
observations. Furthermore, in  existing VS models, the SUSY breaking
vacuum is only metastable.

In HS models, the SUSY breaking sector communicates with the standard
model only via nonrenormalizable interactions suppressed by powers of
$M_P$.  Among these there is a
further dichotomy: the HS may break SUSY and have a stable vacuum in flat
space and only require nonrenormalizable operators to communicate SUSY
breaking to the visible world;
or SUSY breaking may itself arise from Planck scale
nonrenormalizable interactions. The former models, where SUSY breaking
occurs due to renormalizable couplings, we denote RHS, while models
which require nonrenormalizable operators for SUSY breaking are referred
to as NRHS models.
%There is a possible further subdivision
%of this latter class.  The flat space limit of the theory may either
%have a stable supersymmetric vacuum, or a ``vacuum at infinity in field
%space''.  It is possible that gravitational effects stabilize the latter
%class of theories, but, apart from a single toy model discussed at the
%end of section II, we do not consider them further in this paper.

In section II of this paper we study NRHS models.
It is not clear whether there are any explicit examples of
models of this class   which really have stable SUSY
breaking vacua.  The ``racetrack'' models \ref\racetrack{N.V. Krasnikov,
\pl{193}{1987}{37}; L. Dixon, in Proceedings of the A.P.S. DPF Meeting,
Houston, (1990),  L. Dixon, V. Kaplunovsky, J. Louis, and M. Peskin,
unpublished (1990), J.A. Casas \etal, \np{347}{1990}{243}; T. Taylor,
\pl{252}{1990}{59}; B. de Carlos, J.A. Casas and C. Munoz,
\pl{263}{1991}{248}} represent the state of the art.  Our analysis assumes
the existence of such a vacuum, and the phenomenological requirement
(discussed below) that singlet F terms be generated, but no other
specific features of gaugino condensation models.

We first analyze the relation between the scales of breaking of
supersymmetry and R symmetry, and argue that the latter must be
higher than the former if the cosmological constant is to vanish.
In fact, the scaling is precisely right if we assume that the strong
dynamics at the intermediate scale $M_R$ (the SUSY breaking scale is
$M_S \sim
\sqrt{M_R^3 / M_P}$)  spontaneously breaks all R symmetries
down to at most a $Z_2$ ``R-parity''. This is an aesthetic argument in favor
of NRHS models. In theories for which there is
a larger R symmetry down to a scale which   is below
$M_{S}^{2\over 3} M_P^{1\over 3}$, the cosmological constant cannot  be
set to zero even by fine tuning.

We find however that there is a very general cosmological constraint on
NRHS models, which is very hard to satisfy.  We argue  that
such models typically contain particles, including the R-axion,
with masses of order  1~TeV  and
couplings proportional to $1/ M_P$.  Under quite
general assumptions, this field dominates the energy density of
the universe until
 the time when the energy density
is $\sim (10^{-2} \MeV)^4$.  The baryons and radiation that populate the
universe today cannot be produced
until after this epoch. Consequently,
a model incorporating such particles is not compatible with the standard
theory of primordial nucleosynthesis. It is also likely to be
incompatible with a reasonable theory of structure formation in the
universe. Therefore we conclude that our present conceptions of
early cosmology rule out all NRHS models.

The problems we point out in NRHS models are essentially the same as the
problems \ref\polprob{G. D. Coughlan \etal, \pl{131}{1983}{59}} with the
Polonyi field in the earliest hidden sector models  \ref\polonyi{J.
Polonyi, Budapest preprint KFKI-1977-93 (1977); E. Cremmer, S. Ferrara, L.
Girardello, and A. van Proeyen, \np{212}{1983}{413}}. In that context it
has been suggested that the problem can be resolved by raising the
gravitino and scalar masses well above the weak scale
\nref\ceen{E. Cohen, J. Ellis, K. Enqvist and D. V.
Nanopoulos, \pl{161}{1985}{85}}\nref\ellis{J. Ellis, D.V. Nanopoulos,
and M. Quiros, \pl{174}{1986}{176}}\refs{\ceen,\ \ellis}.    In    our
effective Lagrangian analysis, the only mechanism available for raising
the masses of the dangerous scalar fields is to make the dimensionless
coefficients of certain nonrenormalizable couplings very large.  The
general rules of effective field theory suggest that the real meaning of
large dimensionless coefficients is the breakdown of effective field
theory at a scale lower than one had originally assumed.  In other words,
it suggests that the physics responsible for SUSY breaking is operative
at a scale below the Planck mass and that we are not really dealing with a
NRHS model at all.

We conclude Section III by pointing out that string theory is replete
with particles that may suffer from cosmological problems of the type we
have discussed here.  These are the moduli fields, which exist in all
$(2,2)$ string vacua, and all known $(2,0)$ models.  We conclude  that
either one must find string vacuum states without moduli, or that strong
interactions at a scale of $10^{15}~ GeV$ or greater must give mass to
the moduli fields.  Since this is well above the scale of SUSY breaking
(with any conjectured mechanism) we conclude that nonperturbative
effects not associated with SUSY breaking must be an important part of
the stringy description of the world.

After discussing the cosmological problems of NRHS models,
in Section III we show that it is possible to construct viable
RHS models---hidden sector models with flat space
SUSY breaking.  We note however that, assuming a minimal spectrum in the
visible sector, any such theory which also allows gauginos to get weak
scale masses  will have to invoke ``supernaturalness'': certain terms in
the superpotential which are not forbidden by any symmetry must be assumed
absent. String theoretic vacuum states often contain ``stringy
symmetries''\foot{Technically, these are higher spin chiral algebras
on the world sheet.  The most well known of these is the left handed
world sheet N=2 SUSY algebra of $(2,2)$ ground states.} which guarantee
the absence of some superpotential terms at tree level.  Supersymmetric
nonrenormalization theorems then imply that these terms are absent to all
orders in perturbation theory.

  We conclude that if our cosmological constraints cannot be evaded, there
are only two known viable scenarios of dynamical SUSY breaking: the VS
and RHS models.  Both
require the existence of a strongly coupled theory which breaks SUSY in
flat space.  The two mechanisms differ in the medium via which SUSY
breaking is transmitted to the SSM, gravity or gauge interactions.  As a
consequence, the fundamental scale of SUSY breaking in the two classes
of models differs by a factor of order $10^4$ --- $10^7$.
As mentioned above, all known models of dynamical SUSY breaking possess
R-axions. For VS models,
these axions could be seen in laboratory experiments, or could affect
stellar cooling rates and supernova dynamics. This  problem was solved in
\dinelson\ by the  introduction of a second, slightly weaker, gauge
interaction which explicitly breaks the R-symmetry via the anomaly.  For
RHS models, the
R-axion has a decay constant of  just the right size to fit into
the conventional
invisible axion window \ref\invisaxion{Relevant reviews include J.E.
Kim, \prep{149}{1987}{1}; R. Peccei, published in {\it CP Violation},
ed. C. Jarlskog (1988)} and  could be used to solve the strong CP problem.
In the next section we explicitly construct  an example
of a model in which this possibility is realized.  It
requires the imposition of an anomalous U(1) R symmetry, as well as the
absence of several terms which are not forbidden by any symmetry.
Unfortunately,  such a model   seems to be incompatible with fine tuning
the cosmological constant to zero. An alternative (and perhaps more
plausible), scenario for the fate of the R-axion in RHS models, is that
nonrenormalizable Planck scale couplings give the axion a
mass greater than or equal to the weak scale and effectively eliminate its
effects on both cosmology and the strong CP problem.

\newsec{\bf NRHS: Hidden sector models which are supersymmetric in the flat
space limit}

\subsec{The ubiquity of singlets}

A generic feature of HS models is the existence of gauge singlet
superfields which communicate SUSY breaking to the visible world
via interactions suppressed by powers of $M_P$.  The requirement that
such fields be gauge singlets arises from the need  in the SSM to generate both
dimension 3 operators with coefficients of order the weak scale, such as
trilinear scalar couplings and  gaugino and higgsino\foot{We do not
allow for the possibility of putting in weak scale supersymmetric dimension 3
terms such as the mu parameter by hand, but assume they must be generated by
the  same physics which breaks supersymmetry.} masses, as well as  dimension 2
operators  (eg, squark masses), with coefficients of order the weak scale
squared.

Squark masses will always be of the form
$$m_{\tilde q}^2 \sim  {\langle F\rangle^2\over M_P^2}$$
where $\sqrt{\langle F\rangle}$ has dimensions of mass and
is the scale of SUSY breaking in the hidden sector.
In order to explain the gauge hierarchy, one must have
$\langle F\rangle \ltap M_P\times(1\,\TeV)$.
The coefficients of dimension 3 operators,
such as the gluino mass, for example, are of the form
\eqn\mgino{m_{\tilde g} \sim  \left({\langle F\rangle\over M_P}\right)
\left( {\langle F\rangle\over M_P^2}\right)^n\ .}
The gluino mass must be of the same order in $\langle F\rangle/M_P^2$ as
the squark mass, or else it is essentially massless at tree level.  Thus
a large gluino mass
requires $n=0$ in \mgino\  and therefore
arises from a dimension five operator,
\eqn\gauginomass{{\CO(1)\over M_P}\int d^2\Theta\; Z\; W_{\alpha}^2\ ,}
where $W_\alpha$ is the color superfield and $Z$ is the agent of SUSY
breaking. Consequently, $Z$
must carry zero charge under any unbroken symmetry of the theory,
although it may shift under an R-symmetry \foot{The only known
models which escape this conclusion add  extra  chiral superfields to
the visible sector, such as a color octet and some weak triplets, and
require that hypercharge be unified into a nonabelian symmetry at a
scale below $10^{10}$ GeV \ref\gmass{L.J. Hall and L. Randall,
\np{352}{1987}{289}; M. Dine and D. MacIntire,
\physrev{46}{1992}{2601}}.}.

 There is still no conclusive evidence for the
necessity of tree-level dimension 3 operators in the SSM, since there may
remain a small window for a light gluino whose mass is generated
radiatively \ref\gloops{R. Barbieri, L. Girardello, and A. Masiero,
\pl{127}{1983}{429}; R. Barbieri and L. Maiani, \np{243}{1984}{429}}.
However, it should be noted that versions of the SSM
without dimension 3 operators must contain gauge singlets in the low
energy visible sector in order to break an effective $R$ symmetry that
would forbid radiative gluino masses and revive the old $U(1)$ problem
of a light $\eta$ meson.   Therefore {\it any} HS theory of dynamical SUSY
breaking must contain a gauge singlet superfield.
The phenomenological window for a light gluino is small
and a light gluino should either be found or
ruled out in the near future.

 In this paper we will  consider the consequences
of HS models which contain a set of
singlet chiral superfields $z^A$ in the hidden sector.
First we examine NRHS models in which the $z^A$
singlet fields are
coupled to a strongly coupled hidden sector with dynamical mass scale
$M_R$ via nonrenormalizable
interactions of the form
\eqn\lcouple{\delta {\cal L} = c_{A_1 \ldots A_N} {z^{A_1} \over M_P}
\ldots { z^{A_N} \over M_P} {\cal O}^{4}}
where ${\cal O}^{4}$ are operators of dimension $4$ in the
strongly coupled theory.  In the absence of these couplings, the
strongly coupled theory is assumed to have a stable supersymmetric
vacuum.  If such operators exist at dimension 5, then the
SUSY breaking scale in the visible sector will then be of
order $m_{3/ 2} \sim {M_R^3/M_P^2}$---one power of $1/M_P$ arising
from the operator \lcouple, and another from the gravitational
communication of SUSY breaking to the visible sector.
We will also assume that, like the dilaton and moduli superfields in
string theory, these fields have no renormalizable couplings in
the fundamental Planck scale Lagrangian.
It is important to our analysis that the fields $z^A$ are
always small (relative to $M_P$) at the minimum of their effective
potential, which we will {\it assume} to violate SUSY.  Conventional
field theoretic dimensional analysis depends on the assumption that the
vacuum expectation values of fields are much smaller than the scale which
characterizes  irrelevant  interactions.

The only extant examples of models which might belong to the class
defined above, are the ``racetrack'' models of competitive gaugino
condensation \racetrack.  The models are derived from string theory and
the origin of field space is taken to be a point which is natural from
the string theoretic point of view.  It is a distance of order $M_P$
from the minimum.  The use of an effective Lagrangian over such a large
expanse of field space is justified by an appeal to knowledge of the
underlying short distance physics.
Our analysis of these models would begin by
reexpanding the effective Lagrangian around the minimum.  We are not
trying to establish the existence of a SUSY breaking minimum in
a particular model, but rather to demonstrate some very general
consequences of the existence of such a minimum.

We are now in a position to write down the effective Lagrangian for the
scalar fields $z^A$ which results from integrating out the strongly
interacting fields which get mass at the scale $M_R$.  Since the theory
at $M_R$ does not violate SUSY, the effective theory below the scale
$M_R$  has the conventional supergravity form
with effective potential
\eqn\leff{V = e^{{K(z, \bar{z})\over M_P^2}}\( (K^{-1})_B^A \( D_A W D^B
\bar W \)-
{3\over M_P^2}|W|^2 \)}
where
\eqn\coder{ D_A W =  W_A +{K_A\over M_P^2}W\ ,}
and subscripts denote differentiation with respect to $z^A$.
By convention, we
have chosen $z^A = 0$ to be the minimum, and we expand the Lagrangian in
a power series around the origin.  We will choose a Kahler gauge for
which all holomorphic polynomials in $z$ have been shifted from the Kahler
potential $K$ into the superpotential $W$.  The
leading nontrivial term in the Kahler potential,
is of order $z \bar{z}$ with dimensionless
coefficients.  By    linear field redefinitions, we can always
bring it to the form $z^A \bar{z}^A$.  The corrections to this canonical
form may be written
\eqn\kahlerpower{\eqalign{K = &\ha \bar{z}_A z^A + \ha {K_{AB}^C \over
M_P}
 z^A z^B \bar{z}_C\cr
&+ {1\over 3\! M_P^2}
K_{ABC}^D z^A z^B z^C \bar{z}_D + {1\over 8 M_P^2} K_{AB}^{CD}
z^A z^B \bar{z}_C \bar{z}_D + {\rm h.c.} +\ldots\cr}}
The dimensionless coefficients in this expansion are functions of $M_R
/ M_P$ which should   approach finite numbers in the limit $M_R
\ll M_P$.
%(remember that we are describing the coefficients in a
%Wilson-Kadanoff effective action).

Similarly, the effective superpotential has an expansion:
\eqn\potpower{W = M_R^3 \Big(\omega + \omega_A {z^A \over M_P} +
\omega_{AB} {z^A z^B \over M_P^2} + \ldots\Big)}
The power of $M_R^3$ in front of this term reflects our assumption that
strong dynamics at the scale $M_R$ are responsible for generating this
superpotential.  The $M_P$ scaling of terms that depend on the fields
reflects the fact that the singlets couple only through terms of the form
\lcouple .  Again, we expect the dimensionless coefficients to be of
order one.  It is conceivable that some of them are much smaller than
one. For example, our strongly coupled sector might have several scales,
with different symmetries broken at different scales.  If $M_R$ is the
largest scale at which a superpotential is generated, the complications
of the strongly coupled sector would show up as anomalously small
dimensionless coefficients.  Another way to generate small coefficients
is to assume that some of fields $z^A$ couple through operators of
dimension higher than those in \lcouple .  This would make some of the
coefficients positive powers of
 $M_R / M_P$.  On the
other hand, there is no natural way to make the dimensionless
coefficients in this Lagrangian large.  Indeed, the typical meaning of a
large dimensionless coupling in an effective Lagrangian is that we have
made an error in identifying the scale of high energy physics which is
being ignored in the effective field theory. If the high energy scale
is in fact lower than we had imagined, we naturally obtain large
coefficients.  In the present context, this means that the Lagrangian
\lcouple\ is induced by physics below the Planck scale, in contradiction
to our initial hypothesis.

\subsec{R-symmetry and the cosmological constant}

We now impose the constraint that our Lagrangian has a SUSY violating
minimum at $z^A = 0$, with zero cosmological constant.  SUSY violation is
ensured by the fact that $\omega_A \neq 0$.  The constraint that the
cosmological constant is zero is
\eqn\cosmocon{\omega_A \bar{\omega}^A - 3 |\omega |^2 = 0}
Note that this is a cancellation between numbers that are a priori of
order one.  In a more general $N = 1$ supergravity theory, the
cancellation of the cosmological constant implies that the value of the
superpotential at the minimum be related to the SUSY breaking F term by
$ W \sim M_P F\equiv M_PM_S^2$.  A nonzero value for the superpotential breaks
any
R symmetry of the theory larger than $Z_2$.  A natural explanation for this
general order of magnitude relation might be that the theory has an
unbroken R symmetry (perhaps discrete), which is dynamically broken at a
scale \eqn\rscale{M_R \sim \(M_{S}^{2} M_P\)^{1\over 3}\ .}
NRHS models implement this
explanation dynamically if we assume that the strongly coupled hidden
sector at $M_R$
%(note our prescient choice of notation)
is the agent of R-symmetry breaking  (as it is
in models based on gaugino condensation).
On the other hand, {\it if there is an exact R-symmetry which survives to
scales much below \rscale, the cancellation of the cosmological constant
is impossible}.

This observation  is intriguing, but may have
limited relevance for SUSY model building.  Although it rules out models
with exact R symmetries below the scale~\rscale, it does not prevent the
appearance of accidental R symmetries (in particular, the accidental R
symmetries which are the key to all known models of dynamical SUSY
breaking in flat space).  So we obtain only the mildly restrictive new
rule:   R symmetries (other than R-parity) in low energy SUSY models must
be accidental {\it i.e.} they must follow as a consequence of other exact
symmetries plus renormalizability.  We can expect them to be broken by
nonrenormalizable terms scaled by $M_R$ (or perhaps $M_P$).  It is also
impossible to rule out models in which all R symmetries are broken at a
scale higher than \rscale.  For example, many tree level superstring
vacua have unbroken SUSY
and no R symmetries, and the cosmological constant vanishes despite
this.
This may be viewed as a superstring
miracle \ref\dineseib{M. Dine and N. Seiberg,
\np{301}{1988}{357}; \np{306}{1988}{137}} : the constant in the superpotential,
which is not forbidden by any field theoretic symmetry of these vacuum states,
vanishes ``by accident''.  In these vacua, the relation \rscale\ is violated,
but
the cosmological constant might well turn out to be zero.

NRHS models thus enjoy the  distinction of being able to
explain dynamically the relation \rscale,   which, although it by no
means solves the cosmological constant problem, makes the vanishing of the
cosmological constant   a little more natural.
Unfortunately, as we now explain, all NRHS models suffer from a cosmological
problem which makes them unattractive candidates for models of the real
world.

\subsec{Light scalars: a cosmological problem}

The condition that $z^A = 0$ be a stationary point of our action can be
written:
\eqn\stationarity{\bar{\omega}^A \omega_{AB} + \omega_B - K_{AB}^C
\bar{\omega}^A \omega_C =0 }
Note that the noncanonical term in the Kahler potential can be important
in satisfying this condition.  We now want to examine the quadratic
terms in the expansion of the potential around the stationary point.
This task is somewhat simplified by the observation that the prefactor
in the potential can be set equal to one, because the term it multiplies
is already of quadratic order in the $z^A$ (as a consequence of the
conditions \stationarity\ and \cosmocon ).  Note however that the
quadratic terms do depend on the quartic terms $K_{ABC}^D$ and
$K_{AB}^{CD}$
in the Kahler potential.  Since these are as yet unconstrained, we are
free to choose them in such a way that the origin is a stable minimum
with no flat directions.  Thus there are no {\it a priori} arguments
against the existence of NRHS models.
%\foot{as one might have begun to suspect in view of the difficulty of
%constructing them via gaugino condensation.}.
What we cannot do however is to
arbitrarily choose the size of the mass terms for the $z$ fields.  Their masses
are of order ${M_R^3/M_P^2 } \sim m_{3/2}$.  Some of these masses might be
smaller, if the dimensionless coefficients in \kahlerpower\ or \potpower\
are small for the reasons discussed above.  However, for those singlets
which give rise to SSM gaugino masses, there cannot be a symmetry which
prevents all couplings of the form \lcouple.  These will generically
have masses of order the SUSY breaking scale in the visible sector,
which is generally assumed to be of order 100 GeV--1 TeV in order to
eliminate fine tuning problems at the weak scale.

The cosmological consequences  of scalar fields with    mass
$m$ and gravitational strength interactions have been analyzed before
\nref\ross{G. German and G.G. Ross, \pl{172}{1986}{305}}\nref\berto{O.
Bertolami, \pl{209}{1988}{277}}
\refs{\polprob,\ \ellis,\ \ross,\  \berto }, and we briefly
review the results here.  In the early universe, typical thermal values
of the scalar field strength are far from the origin (which we take to be
the low temperature minimum of the potential), and so there is a Bose
condensate. Such a condensate cannot be eliminated by
inflation \nref\inflation{A.H.
Guth, \physrev{23}{1981}{347}; A. Linde, \pl{108B}{1982}{389}; A.
Albrecht and P. Steinhardt,
\prl{48}{1982}{1220}
}\nref\linde{A.S. Goncharev, A. D. Linde, and M. I. Vyssotsky,
\pl{147}{1984}{279}}\refs{\inflation,\ \linde,\ \ross}\foot{unless the Hubble
constant during the inflationary period is smaller than the mass of the scalar
field. However,  with such late inflation it is not possible in a natural
model to reheat after inflation to a temperature above $\sqrt{m^3/M_P}$,
which is below the temperature needed for nucleosynthesis if $m<1$
TeV.}.        When the temperature cools below  $T \sim \sqrt{m M_P}$ the
spatially averaged value of the scalar field oscillates about the
minimum, and so we have a gas of  cold bosons. Assuming that the inital
expectation value of the field  is of order $M_P$, if   $m$ is
greater than $10^{-28} \eV$, these bosons contribute too much energy
density to the mass of the universe to be consistent with standard
cosmology, unless their lifetime $\tau$ is sufficiently short so that
they decay before nucleosynthesis.  However for bosons with
gravitational strength interactions,   $\tau$ must be greater than
$\CO(M_P^2/m^3)$, and once they have decayed the temperature $T_R$ of the
universe is $\CO(\sqrt{M_P/\tau})\ltap\sqrt{m^3/M_P}$. Standard
nucleosynthesis requires that $T_R$ be larger than $\CO(1)$ MeV, and so
that the mass $m$ be larger than $\CO(3) \times 10^4 $ GeV. Since, in the
models of interest, both $m$ and the weak scale are naturally less than or
of order the gravitino mass,  conventional nucleosynthesis requires a
supersymmetry breaking scale which is uncomfortably large for a natural
resolution of the gauge hierarchy problem. Furthermore, the scalar decays
generate a tremendous amount of entropy, which would dilute any initial
baryon abundance to well below what is observed. Unless the decays also
lead to baryon number production\foot{For example the scalar decays
could proceed via baryon number and CP violating interactions.}, then we
must demand that $T_R$ is sufficiently large  for baryogenesis. For
instance, in order to allow for baryogenesis during the weak phase
transition \ref\wsb{For a recent review, see A.G. Cohen, D.B. Kaplan and
A.E. Nelson, UCSD preprint UCSD-PTH-93-02 (1993), to be published in Ann.
Rev. of Nucl. and Part. Phys., vol. 43} $m$ and $m_{3/2}$ must be greater
than $\CO(10^8) \GeV$. We conclude that the cosmological constraints on
NRHS models make these theories much less attractive.

\subsec{Exotica}

In the above discussion we assumed both that the HS had a stable
supersymmetric vacuum in the $M_P\rightarrow\infty$ limit, and that the
$z^A$ fields had no renormalizable couplings.  We now briefly consider
what happens if one relaxes those two conditions.

First we offer a toy example of a model in which gravitational effects could be
responsible for supersymmetry breaking, but for which there is no stable
vacuum as $M_P\rightarrow\infty$---the example is simply massless QCD, with
fewer flavors than colors. In flat space, there is a dynamically generated
superpotential  of the form \sqcd \eqn\qcdpot{W_{\rm eff}=
\CO(1)\Lambda^{3n_c-n_f\over n_c-n_f}\({   1\over \det_{ij}\bar
q_i^a q_j^a}\)^{1\over n_c-n_f}\  ,}where $i,j$ are flavor indices, $a$ is
a color index, $\Lambda$ is the QCD scale and $n_c$ and $n_f$ are  the
number of colors and flavors respectively. This  superpotential drives
the squark vevs to infinity and the theory has no ground state. However
if one naively includes the effects of supergravity by using the effective
potential \leff, assuming a Kahler potential of form \kahlerpower, then
one finds that the vevs could be stabilized at a value of order $M_P$
(where, however, an effective field theory analysis breaks down) and the
supersymmetry breaking scale is then\eqn\susyscale{M_S\sim\(\Lambda^{3
n_c-2 n_f}\over M_P^{n_c}\)^{ 1\over 2(n_c-n_f)}\ .}    Although because
of the large field strengths there is no reliable way of computing whether
or not gravitational effects actually stabilize this theory and break
supersymmetry,  one can see using dimensional analysis that even  if this
is the case, there will be a light scalar particle of mass $M_S^2/M_P\sim
m_{3/2}$, with gravitational strength couplings to other light fields. Because
this theory has a spontaneously broken  exact  U(1) R symmetry, the R-axion is
a
massless Goldstone boson, with  decay constant of order $M_P$.

Next we consider models in which the $z^A$ fields have renormalizable
couplings to other fields, and which might also produce a SUSY
breaking scale vanishing as $M_P\rightarrow\infty$.
The generic example is of the form
\eqn\renhs{c_A z^A R \bar{R}}
where $R$ is some representation of the strongly coupled hidden sector
gauge group, or of the visible sector gauge group.   We assume this strongly
coupled theory has a stable, supersymmetric vacuum to fit in the NRHS
category of models.  First consider the $M_P\rightarrow\infty$ limit.
On intregrating out the strongly interacting fields
$R$, the scale $M_R$ will generically be induced in the effective Kahler
potential for the singlets $z^A$.  {\it If at the scale $M_R$
a direct effective superpotential for the fields $z^A$ is not generated}, then
the scalar potential is flat and only
Planck scale effects could possibly break SUSY. For instance,
when Planck scale effects are included and the scalar potential \leff\
is expanded in inverse powers of $M_P$,
 the effective
superpotential for the singlets below the scale $M_R$ of the
strongly coupled physics could be of the form \eqn\singpot{
{M_R^3\over M_P} z_A\ ,}which would give SUSY breaking at a scale
$\sqrt{M_R^3/M_P}$.  The resulting theory has a light particle,
but it could receive a radiatively generated mass   of order ${M_R^2
/ M_P}
 \sim (m_{3/ 2}^2 M_P)^{1/3}$ rather than
$m_{3/ 2}$ (which we have shown is
generic in NRHS models), and it can have   couplings which are
suppressed by $M_R$ rather than $M_P$.  The origin of this relatively
large contribution to the scalar masses is a noncanonical term
in the Kahler potential of the form
\eqn\noncan{K_{AB}^C z^A z^B {\bar{z}_C \over M_R}}
induced by the strong interaction between the scalars and the hidden sector.
We regard this class of models as
somewhat special, since if we assume, in accord with all existing
evidence, that nonperturbative effects generate any superpotential
consistent with the symmetries of the model, then it is very hard to
satisfy the condition italicized above.  Indeed, we have so far been
unable to find any SUSY breaking models in this class in which all mass
scales below the Planck mass are generated dynamically, and which is
natural in the sense that all terms consistent with the symmetries are
included.  Thus, although this class of NRHS models may evade our
cosmological bound, it does not seem attractive.  It is concievable that the
class
is in fact empty.

\subsec{Implications for string theory}

Brustein and Steinhardt have recently pointed out
cosmological maladies of superstring inspired models even more severe
than those we have
discussed here  \ref\brst{R. Brustein and P.
Steinhardt, \pl{302}{1993}{196}}.  These difficulties
involve specific properties of  the dilaton superfield and are a
cosmological version of the Dine-Seiberg problem \ref\dsp{M. Dine, N.
Seiberg, \pl{162}{1985}{299}}.  The constraints on NRHS models that we have
discovered are much more general.  They do not assume that the dilaton
has anything to do with SUSY breaking, and apply equally well to NRHS
models which have nothing to do with string theory.

Note that if, as suggested by our constraints, the mechanism which fixes
the dilaton vacuum expectation value does not break SUSY,  the
Brustein-Steinhardt problem  is more easily resolved \foot{ An
alternative solution based on the ideas of chaotic inflation has been
suggested by Shenker \ref\shenker{S. Shenker, private communication}.}.
For instance,  the dynamics which stabilizes the vevs of the dilaton
and  other moduli could be strong at a scale not far below $M_P$, leading
to a large barrier between the desired ground state and the disastrous
vacuum at infinite dilaton VEV. For instance if the moduli are stabilized
by strong dynamics at a scale of $\sim10^{16}$ GeV,   they could give rise
to  natural inflation at a
scale which gives the desired primordial density fluctuations
\ref\natinf{K. Freese, J.A. Frieman, and A.V. Olinto,
\prl{65}{1990}{3233}}, with subsequent reheating to $10^{10}$ GeV. In any
case, for the universe to be at least at a temperature of 1 MeV (as
necessary for nucleosynthesis) after the moduli decay, they should be
stabilized by dynamics above  $10^{14}$ GeV.  Thus, there appears to be
a general discrepancy between the scale of supersymmetry breaking required by
phenomenology and the mass scale for moduli fields required to avoid the
cosmological difficulties of light weakly coupled bosons.  It appears
that a consistent string theoretic description of nature requires
nonperturbative dynamics at a scale higher than than of SUSY breaking, or
the existence of string vacua without moduli.

Finally, let us add to our catalogue of mechanisms for dynamical SUSY
breaking the still mysterious proposal of ``stringy nonperturbative
breaking''\ref\shenker{S.H. Shenker,  talk given
  at the Cargese Workshop on Random Surfaces, Quantum Gravity
and Strings, Cargese, France,   (1990),
Published in the proceedings.}  Since SUSY
breaking must occur well below the Planck scale, these effects must be
encodable in a superpotential in a supersymmetric low energy effective
lagrangian just above the SUSY breaking  scale.  Since the physical states
contributing to these effects all have masses of order the string scale
\shenker, we expect a contribution to the superpotential of the form
\eqn\nonpert{\epsilon M_S^3 W\({z_i\over M_S}\)}
where $z_i$ are some chiral superfields, and $M_S$ is the string scale,
which is only slightly smaller than the Planck mass. $\epsilon$ is the
small parameter ($\sim e^{-{1\over g}}$) which identifies this as a
nonperturbative effect in string theory.
This is essentially the same kind
of potential that we analyzed above
and it leads to the same kind of problems.  Our ignorance of the
nature (and even the existence) of stringy nonperturbative SUSY breaking
is so great that we should take this argument with a grain of salt,
but one should certainly worry that this hypothetical mechanism for SUSY
breaking may encounter cosmological problems.

\newsec{\bf RHS: Hidden sector models which dynamically break SUSY in the
flat space limit}

Having discussed  NRHS  models in which SUSY is restored in the
$M_P\rightarrow\infty$ limit, we turn to RHS models where gravity serves
only to communicate SUSY breaking. We find that while RHS models can easily
evade the cosmological problems found for NRHS scenarios, it is difficult to
give the visible gauginos a weak scale mass even with the introduction of
gauge singlets. We do explicitly construct a model, where there is a
hidden sector contribution to gaugino masses which is suppressed relative to
the squark and slepton masses by a loop factor.

 It has been known for some time that dynamical supersymmetry
breaking  is possible  in a renormalizable field theory in flat space
\refs{\susybreak,\ \ads}, and that it is difficult to incorporate this
mechanism into a realistic model.   An example of flat space dynamical
supersymmetry breaking in the visible sector is given in \dinelson. The main
problem with RHS models is that   the   gauginos
in the visible sector get only a very small  mass \ads. One alternative, with
greatly expanded visible sector, is described in ref. \gmass; here we will
assume that the visible sector has minimal field content. Another alternative
is that the gauginos gain mass only from loops involving visible sector
particles (and from electroweak symmetry breaking) \gloops, in which case
evidence for light gluinos ($\sim 500$ MeV) or charginos ($\sim 50$ GeV)
could appear soon. If larger gaugino masses are required, then the hidden
sector must contain a gauge singlet as discussed in section~2, with a
coupling to the visible gauge fields   as in eq.~\gauginomass. This singlet
must also couple to the hidden supersymmetry breaking sector with {\it
renormalizable} couplings, since in order to give the gauginos a weak scale
mass its F component must be of order the supersymmetry breaking scale
squared. However existing models of  flat space dynamical supersymmetry
breaking only contain matter superfields in {\it chiral} representations of
the gauge group, and so can couple to a gauge singlet only via terms of
dimension five or more.  Thus the simplest possible strategy, which is  to
take for the hidden sector     a    model of dynamical supersymmetry breaking
with matter fields in purely chiral representations,  leads to very light
gauginos in the visible sector.

In order to give gauginos   a  larger mass, we need to find a model of flat
space dynamical supersymmetry breaking in which some matter superfields are
in real representations of the gauge group, so that they can couple to a
gauge singlet via renormalizable interactions. Such   models are not
impossible to find. An example   was given in \dinelson, but this model has
the drawback of having a supersymmetric minimum at infinity in field space,
as well as a local supersymmetry breaking minimum.  Here we will give an
example of such a model which probably has a unique supersymmetry breaking
ground state.

Many dynamical supersymmetry breaking models contain global symmetries, which
we can weakly gauge without affecting the supersymmetry breaking dynamics. Our
strategy is to take such a model, and  to then add superfields in real
representations of this new weak gauge symmetry, which can also couple to
a the gauge singlet. For example, we can take SU(7) for the
``supercolor'' gauge group, whose dynamics are responsible for breaking
supersymmetry, with matter superfields in the representation ${ {\bf
21}\oplus 3\ {\bf \bar 7} }$. It has been argued that this theory
breaks supersymmetry and has a stable ground state \ads\ if there is a
nonzero superpotential ${W=\lambda_1{\bf 21}_{a b}{\bf \bar 7}^a_1{
\bf \bar
7}^b_2  }$, where $a,b $ are SU(7) indices. An SU(2) global symmetry is
preserved by this superpotential, which we  weakly gauge, requiring that
we add an odd number of SU(2) doublets to the theory to cancel the SU(2)
anomaly. The minimal set of matter fields, which will allow us to couple a
gauge singlet  to the supersymmetry breaking sector via renormalizable
interactions, transforms under SU(7)$\otimes$SU(2) as  \eqn\mattrep{{\bf
(21,1)}\oplus  {\bf (\bar 7,2)}\oplus   {\bf (\bar 7,1)}\oplus 3\  {\bf
(1 ,2)}\oplus {\bf (1,1)}\ . } For a superpotential we take
\eqn\superpot{W_{\rm hidden}=\lambda_1\epsilon^{i j}{\bf (21,1)}_{a
b}{\bf (\bar 7,2)}^a_i {\bf (\bar 7,2)}^b_j   +
\lambda_2\epsilon^{i j}{{\bf (1
,2)}_1}_i{{\bf (1 ,2)}_2}_j{\bf (1 ,1)} \ ,} where $i,j$ are SU(2)
indices.  Unfortunately the ground state of this model is strongly
coupled \ads, and we cannot reliably compute its features. However it is
at least plausible that supersymmetry breaking induces a negative mass
squared for the scalar components of the ${\bf (1 ,2)}$ representations,
and induces expectation values for them. We will also assume that
nonsupersymmetric quartic terms are induced, which stablize the expectation
values at finite values. Then the F term for the gauge singlet will be
nonzero.

In order to communicate supersymmetry breaking to the gauginos in
the visible sector, we add to the Lagrangian a term  \gauginomass, where
$Z$ is now the gauge singlet field ${\bf (1 ,1)}$. Supergravitational
interactions will induce masses  for the scalar fields of the
visible sector which are of order the gravitino mass. Since the F term
for the singlet, which is responsible for the visible gaugino masses, is
only induced at one loop by the weak SU(2) gauge coupling,  the gaugino
masses will be much smaller than the supersymmetry breaking scalar
masses, by a factor of order $\alpha_{{}_{\rm SU(2)}}/\pi$. Therefore this
model  has rather light gauginos, although the hidden sector
contribution to their mass is larger than the contribution arising from
loops in the visible sector.

The hidden sector of this model has an exact nonanomalous U(1) R
symmetry,  which is spontaneously broken at the supersymmetry breaking
scale, and hence  has an R-axion. It is possible to arrange the couplings
to the visible sector such that the R symmetry is broken only by a color
anomaly, and  the R-axion is just the usual axion  which solves the strong
CP problem. Remarkably, a gravitino mass of order the weak scale results
from a supersymmetry breaking scale of $\sim 10^{11}$ GeV, which fits
precisely into the allowed window for the invisible axion decay constant.
 The singlet field $Z$ must have R charge zero, so that the coupling
\gauginomass\ is allowed, and we take the Higgs fields of the visible
sector to also have R charge zero, which forbids a direct
supersymmetric mass term. If we then take the Kahler potential to include
terms \eqn\kpot{ \int d^4\theta\; {\CO(1)\over M_P}Z^* H_1 H_2 +
{\CO(1)\over M_P^2}H_1 H_2\(\sum_n s_n s^*_n\)\ ,}  where the $s_n$ are
generic hidden sector superfields, we will induce a small higgsino mass
term (a ``$\mu$ term'') of order the gaugino mass, as well as a larger
supersymmetry breaking mass term for the scalar components of the Higgs.
Thus this model has several desirable features: all mass scales
other than the Planck mass are generated dynamically,   the
``$\mu$-problem'' is solved by the mechanism of Guidice and Masiero
\ref\guidice{G.F. Giudice and A. Masiero, \pl{206}{1988}{480}}, and there
is no   strong CP problem. One still must worry about electric dipole
moments which could be induced by CP violation in the terms in eq.~\kpot,
(in the minimal supersymmetric standard model the relative phase between
these two terms corresponds to the phase
 which is usually called $\phi_B$, which induces quark and lepton dipole
moments at one loop). However, these dipole moments are suppressed by the
small size of the gaugino and higgsino mass terms relative to the squark and
slepton masses, and can  be within experimental bounds for  phases of order
one.

Unfortunately there are several troubling naturalness problems with
any model of this type. For one thing the superpotential \superpot\ is not
the most general allowed by the symmetries. This is a general   problem,
as can be seen by the following argument. Because of the coupling
\gauginomass, the singlet field cannot carry any non-zero  charges except
a shift symmetry $Z\rightarrow Z+i\ {\rm constant}$, and we also need the
singlet to have a renormalizable coupling to the supersymmetry breaking
sector, \eg  \eqn\zrrb{Z \bar R R\ ,} where $R$ and $\bar R$ are
superfields coupled to the supersymmetry breaking sector.  There is no
symmetry which allows the couplings \gauginomass\ and \zrrb\ which does
not also allow a coupling $\bar R R$,  and so   $R$ and $\bar
R$, which are responsible for communicating supersymmetry breaking to the
singlet, naturally would have mass of order $M_P$, which would greatly
suppress their ability to communicate supersymmetry breaking to the
singlet. It is therefore necessary to invoke ``supernaturalness'', \ie\
the fact that in supersymmetric theories it is technically natural to omit
arbitrary terms from the superpotential which are allowed by all symmetries of
a theory, and this does not destroy the renormalizability of the theory. For
instance, such missing terms often arise in string theories. Still, we find it
unsatisfying to have to rely on unknown  Planck scale physics to explain
the absence of undesirable terms. An even worse feature of this model is
that the R symmetry which solves the strong CP problem is imposed by hand,
and is not an accidental consequence of other symmetries of the theory.
In the previous section we argued that we cannot require any R symmetries
(other than accidental ones) to be unbroken below a scale $M_R\sim
M_S^{2\over 3} M_P^{1\over 3}$, or it will be impossible to tune the
cosmological constant to zero. Of course, since one has to rely on
supernaturalness anyway to suppress unwanted terms, one need not invoke
any R symmetry, but then there is no reason not to expect the axion to
gain mass from Planck scale physics of order $M_S^{n+1}/M_P^n$, \ie\ the weak
scale for $n=1$. Unless $n$ is larger than 6, such an axion   will play no
role in solving the strong CP problem although it always either decays
quickly enough or contributes a small enough mass density to the universe to
escape  all cosmological problems.

In summary, dynamical supersymmetry breaking in flat space is a viable
option for hidden sector models, although one which will typically lead to
light
gauginos in the visible sector. The simplest possibility is that the hidden
sector model has purely chiral matter field content, and so   visible gaugino
masses only arise from loops in the visible sector and are very small (e.g.
$\sim
500$ MeV for the gluino). Somewhat larger  gaugino masses are possible if one
complicates the hidden sector to allow for a gauge singlet which has
renormalizable
couplings with the supersymmetry breaking sector. Any such model requires
the omission of some terms in the effective Lagrangian which are not
forbidden by any symmetry. In  explicit models which we have
constructed,   the gaugino masses are still suppressed relative to the
scalar masses by a weak hidden sector gauge coupling. Such models
generically contain an R-axion with a decay constant equal
to the  SUSY breaking scale--this could be identified with the invisible
axion, and in general presents no cosmological problems.

\newsec{\bf Summary}

There are many possible scenarios for solving the gauge hierarchy
problem via dynamical supersymmetry breaking. Here we have shown that
models in which Planck scale effects   give rise to dynamical
supersymmetry breaking   typically contain light (less than 1 TeV) scalars
with gravitational strength couplings.   For instance all string inspired
models, where supersymmetry
breaking is driven by the same dynamics which stabilizes the moduli,
have such light, weakly coupled states. These are difficult to fit into an
acceptable cosmological scenario even when inflation is taken into
account. More promising are models in which supersymmetry is dynamically
broken in flat space. Hidden sector   models  with flatspace
supersymmetry breaking typically produce very small masses for the visible
gauginos. Increasing the gaugino masses in
such models requires either a greatly expanded visible sector, or the
introduction of a singlet in the invisible sector with renormalizable
couplings  and the omission of some terms which are not forbidden by any
symmetry. Furthermore the gaugino masses are still suppressed compared with
squark and slepton masses. Thus in all known viable hidden sector models,
the gauginos could be discovered soon.  Models in which supersymmetry is
broken dynamically in the visible sector can be constructed which have no
difficulties in producing weak scale gaugino masses, which explain the
absence of flavor changing neutral currents in a simple way, and  which
present no phenomenological or cosmological difficulties. We conclude that
the latter class of models deserves further study.

{\bf Acknowledgements:} We are grateful to Nathan Seiberg for many useful
comments
and critcisms. A.N. and D.K. thank the theory goups at Rutgers and at CERN for
their hospitality.
\listrefs

   \end